\documentclass[onecolumn,showpacs,preprintnumbers,amsmath,amssymb]{revtex4}

\usepackage{epsf}
\usepackage{graphicx}  
\usepackage{dcolumn}   
\usepackage{bm}        

\newcommand{\be}{\begin{equation}}
\newcommand{\en}{\end{equation}}
\newcommand{\bea}{\begin{eqnarray}}
\newcommand{\ena}{\end{eqnarray}}

\newcommand{\p}{\partial}
\begin{document}


\title{Some properties of a new class of plane symmetric solution}

 \author{Hongsheng Zhang\footnote{Electronic address: hongsheng@kasi.re.kr} }
 \affiliation{\footnotesize
 Korea Astronomy and Space Science Institute,
  Daejeon 305-348, Korea }
 \affiliation{\footnotesize Department of Astronomy, Beijing Normal University,
Beijing 100875, China}
 \author{Hyerim Noh\footnote{Electronic address: hr@kasi.re.kr} }
 \affiliation{\footnotesize
 Korea Astronomy and Space Science Institute,
  Daejeon 305-348, Korea }
 \date{ \today}

\begin{abstract}
   A new class of static plane symmetric solution of Einstein field equation, which is judged
    as the source of Taub solution, was presented in our previous
    work. In this letter the properties of geodesics of this solution are explored. It is
   found that this solution can be an appropriate simulation to the field of a uniformly
   accelerated observer in Newton mechanics. The
   essence of the source is investigated. A phantom with dust
   and photon is suggested as the substance of the source matter.

\end{abstract}


\maketitle

\section{Introduction}
 We successfully found a new class of static plane symmetric solution of Einstein field
 equation sourced by a perfect fluid in \cite{self}. A very general
 4-parameter solution was presented and then a reduced 2-parameter
 family was studied in detail \cite{self}.

  This solution  puzzles out
 a long standing problem: the source of Taub solution \cite{taub}.
 A no-go result is shown in \cite{taub2}, which says
  that a perfect fluid cannot bound a vacuum in a
  space with plane symmetry unless the boundary condition of the
  continuity of the derivatives of the metric tensor is violated.
  This means that there does not exist a matter source which can
  perfectly match to vacuum Taub solution. A requirement is imposed in the proof of this
  no-go theorem,  that is,  the pressure
 is positive throughout the slab for the solution. However, it is shown that
 generally speaking
  any singularity free source with reflective symmetry for plane symmetric vacuum space
  must violate dominant energy condition (DEC)\cite{dec}. In fact, we find that DEC is always
 violated in the solution in \cite{self}, no matter what values the parameters are taken. Further,
  we have found that there is a configuration in
  the class of \cite{self} which can perfectly
  match to vacuum Taub solution, and  it is naturally to be identified as the
  source of Taub solution.

  To reveal more physical significance of the solution, we need to
  study the motions of massive particles (in the following context called particles) and
  massless particles (in the following context called photons). We will study the geodesics of particles and
  photons
  respectively. Although the
  energy-momentum form of the solution is obtained, we have no idea
  what it really is. We will study the
  possibilities for various candidates, including dust, photon, scalar, and phantom.

 This letter is organized as follows: In the next section we
 present a brief review of the 2-parameter family in \cite{self} as our starting
 point for further discussions. In section III we investigate the geodesics in this space.
 In section IV we study the substance of the source. A
    summary and some discussions are presented in section VI.

 \section{the solution}
  The plane symmetric solution sourced by perfect fluid,
  which is presented in \cite{self}, reads,
  \be
  ds^2=-e^{2az}dt^2+dz^2+e^{2(az+be^{az})}(dx^2+dy^2).
  \label{metric}
  \en
  There are  two free parameters $a, b$ in the above solution.
 With the above metric, $\rho(z)$ and $p(z)$ are respectively given by
  \bea
 \rho(z)&=&-a^2(3+8be^{az}+3b^2e^{2az}),
 \label{rho} \\
 p(z)&=&a^2(3+4be^{az}+b^2e^{2az}),
 \label{p}
 \ena
 where we have set $8\pi G\equiv 1$, and $G$, as usual, denotes the Newton gravitational
 constant. (\ref{rho}) and (\ref{p}) determine an implicit function $p=p(\rho)$. It can be
  treated as the most general form of the source of plane symmetric space in \cite{exactsolution}.

 If we require it to match to a vacuum space, $p$ should vanish naturally  at some
 distance $z_0$ from the ground $z=0$,
 \be
 p|_{z=z_0}=0.
 \en
 Under the conditions
 \bea
  b&=&-3e^{-az_0},\\
  az_0&=&-1/3,
 \ena
  the above solution can perfectly match to Taub solution (Here, we present a minor
  correction to \cite{self}, which does not change any conclusion of \cite{self}).
   As we have shown in \cite{self},
  the metric (\ref{metric}) degenerates to AdS (anti-de Sitter),
  \be
  ds^2=dz^2+e^{2az}(-dt^2+d{x}^2+d{y}^2),
  \label{ads}
  \en
  when $b$ goes to zero.

 \section{Geodesics}
  Free falling is natural state of a particle. Mathematically, its orbit is time-like or null geodesics.
   In fact, any test particle with the same initial position and velocity goes along the same
   orbit: It
   is just the (weak) equivalence principle.  Fundamentally, we can derive the geodesics by the self-parallel
  equations,
  \be
  \frac{d^2x^\mu}{d\tau^2}+\Gamma^{\mu}_{\alpha\beta}
  \frac{dx^\alpha}{d\tau} \frac{dx^\beta}{d\tau}=0,
  \label{geodesic}
  \en
  where $x^{\mu}=(t, z, x, y)$, $\Gamma$ denotes the  affine connection
   of metric (\ref{metric}), $\tau$ is an affine parameter along the
   geodesic. We can calculate the affine connections routinely and
   then solve the four coupled second order equations. Generally speaking it is
   not an easy work.

   Alternatively, we advance a different way by applying the
   properties of Killing vectors. The velocity normalization imposes
   the constraint along a curve,
   \be
   g(\frac{\p}{\p \tau},\frac{\p}{\p
   \tau})=g_{tt}\left(\frac{dt}{d\tau}\right)^2 +
   g_{zz}\left(\frac{dz}{d\tau}\right)^2+
   g_{xx}\left(\frac{dx}{d\tau}\right)^2
   +g_{yy}\left(\frac{dy}{d\tau}\right)^2=\eta,
   \label{normalization}
   \en
   where $g$ is the metric tensor whose components are given by (\ref{metric}),
    $\eta=-1, 0, 1$ for time-like, null, space-like curve,
   respectively. For a time-like curve, $\tau$ denotes the proper
   time; For a null curve, $\tau$ represents an affine parameter;
   and for a space-like curve, $\tau$ stands for the length
   parameter. The inner product of a Killing vector $\xi$ and the tangent
   vector $T=\frac{\p}{\p \tau}$ of a geodesics is a constant along the geodesic. This
   result can be proved as follows,
   \be
   \nabla_{T} (g(T,\xi))=g(\nabla_{T} T, \xi)+g(T, \nabla_{T}\xi)=0,
   \en
 where $\nabla$ is the derivative operator consisting with the
 metric $g$. The three Killing vectors $\frac{\p} {\p t}$, $\frac{\p}{\p x}$, and $\frac{\p} {\p y}$
 yield
 \be
  g(T,\frac{\p} {\p t})=g_{tt}\frac{dt}{d\tau}=-E,
  \label{E}
  \en
   \be
  g(T,\frac{\p} {\p x})=g_{xx}\frac{dx}{d\tau}=P_1,
  \label{P1}
  \en
  \be
  g(T,\frac{\p} {\p y})=g_{yy}\frac{dy}{d\tau}=P_2,
  \label{P2}
  \en
  respectively. Here, $E,~P_1, P_2$ are three constants. Physically,
  for a time-like geodesic  $E$ denotes the energy of a unit mass particle moving along it, $P_1, P_2$ are the momentum
   of unit mass particle along $x$ and $y$ direction, respectively. From now
   on we only consider time-like and null curves since the particles of realistic
   matter can run only  along such curves.

   First we consider the
   motion along $z$ direction, that is $x=$constant, $y=$constant.
   In this case we should solve the equation set
   (\ref{normalization}) and (\ref{E}). Note that the geodesics along $z$ direction will be the same
   as ones
   of an AdS, since the equations to determine it are exactly the same as that of AdS,
   say,  (\ref{normalization}) and (\ref{E}). The reason is that the AdS metric (\ref{ads})
   permits the Killing vector $\frac{\partial}{\partial t}$ and the tangent of the geodesic also needs to
   obey the normalization equation (\ref{normalization}), where $g_{tt}$ and $g_{zz}$ are the same with
   (\ref{metric}).   For time-like geodesic, the
   result reads
   \be
   z=\frac{1}{a}\left[\ln E+\ln\mid\sin(a\tau +c_1)\mid\right],
   \label{ztau}
   \en
   \be
   t=-\frac{1}{aE}\cot(a\tau+c_1)+c_2,
   \label{ttau}
   \en
  where $c_1,~c_2$ are integration constants. It is clear there is a maximum
  value of $z$.  For such an observer, the bigger value of $E$, the farther it can go.
  There is an apparent problem for this observer.
  $z$ and $t$ will be divergent when $a\tau+c_1\to n\pi$, where
  $n$ is an integer. This indicates that $z, t$ are not suitable
  coordinates at the neighborhood of $a\tau+c_1= n\pi$. The case is
  very similar to the case of Schwarzschild coordinate $t$. The
  Schwarzschild coordinate $t$ of a free falling observer will be divergent
  when it goes to the event horizon. To find a well-posed coordinate
  is not very difficult. For example, we set a new coordinate $I$ to
  replace $t$,
  \be
  I=\frac{2(c_2-t)}{1+(t-c_2)^2}=aE\sin(2(\tau+c_1)),
  \en
  and a new coordinate $J$ to replace $z$,
  \be
  J=ae^{z}=E|\sin(a\tau +c_1)|.
  \en
  It is easy to see that for any $\tau$, both $I$ and $J$ keep finite.

   Now we consider its 3-velocity measured by $t$. The 3-velocity is
   defined as
   \be
   \vec{v}=\frac{dx^i}{dt}\frac{\p}{\p x^i},
   \en
   where $i$ runs from 1 to 3. Here only the velocity in $z$
   direction does not vanish, whose magnitude reads,
   \be
   v=[g_{zz}(\vec{v},\vec{v})]^{1/2}=\frac{dz}{d\tau}(\frac{dt}{d\tau})^{-1}.
   \en
    By using (\ref{ztau}) and (\ref{ttau}), we arrive at,
    \be
    v=\frac{a^2E^2(c_2-t)}{1+(c_2-t)^2a^2E^2}.
    \en
    Expanding $v$ about $t=c_2$, we obtain,
    \be
    v=-a^2E^2(t-c_2)+a^4E^4(t-c_2)^3+{\cal O} \left((t_2-c)^5\right).
    \label{exp}
    \en
    In the region around $t=c_2$, constant acceleration is a fairly
    good approximation since there is no $(t-c_2)^2$ term in the
    expansion series. In (\ref{exp}), $v$ is independent of $z, x, y$.
    In such a sense, the global effect of gravitation, not only the
    local effect, can be simulated by a field of a uniformly
    accelerating observer. One may doubt that this is a physical
    conclusion since a coordinate system can be rather arbitrary, that is,
    if we change a coordinate system whether this conclusion remains valid.
    We present a brief analysis of this point.
    $\frac{\p}{\p t}$ is the time-like Killing vector which denotes that the metric
    is static. Hence the coordinate $t$, as its integral curve, is
    unique (up to a constant factor). Similarly, the coordinates $x,
    y$ are unique (up to constant factors). The coordinate $z$ is
    orthogonal to all of them in the 4-dimensional space-time with $g_{zz}=1$. Hence
    it is unique too. So our conclusion is physical. A research on plane symmetric solutions in order to find the
   best simulation to general relativity
    of the Newtonian infinite plane was presented in \cite{jone}.
    For some interesting researches on the equivalence between a uniformly accelerating reference frame and
    the gravitational field, see \cite{equi}.

    For null curves, $\eta=0$, we reach,
    \be
     z=\frac{1}{a}\ln (aE\tau+c_3),
     \en
     \be
     t=-\frac{1}{a^2E^2\tau+ac_3}+c_4,
     \en
      where $c_3$, $c_4$ are integration constants.

      For the most general case in which $x$ and $y$ are not constants, we
      have to solve the associated equation set
      (\ref{normalization}, \ref{E}, \ref{P1}, \ref{P2}). In this case a particle goes along
      a different geodesic since $g_{xx}$ and $g_{yy}$ in (\ref{metric}) become different from that of (\ref{ads}).
      We can derive some physical
      conclusions directly from the equations without analytical solutions. Since the space
      is inhomogeneous in $z$ direction, the geodesic along which $z=$constant does not exist.
      So any initial particles moving in $x$ or $y$ direction also must move  in $z$ direction.
      However, a test particle will
      go through the same proper distance in the same proper time
      along $x$ or $y$ direction in such motions, that is, so it moves uniformly in the $x-y$ plane.
       For example, along $x$ direction
      \be
      \frac{\delta \int_{x_0}^{x}
      \sqrt{g_{xx}d\tilde{x}^2}}{\delta \tau}=P_1,
      \en
      where we have used (\ref{P1}). For the case of null geodesics,
      we have the same conclusion, but we need to change the ``proper
      time" to ``affine parameter".

       For a time-like geodesic, it is
      very difficult to find an analytical solution in the case of $x$ or
      $y$ is not constant. For a null geodesic, we find an
      analytical solution in integral form for $y=$constant. In fact
      it is the most general case because the plane symmetry exists
      on the $x-y$ plane. It is a solution of equation set  (\ref{normalization}, \ref{E},
      \ref{P1}) with $\eta=0$ and $y=$constant, which can be written
      as follows,
     \be
     t=\int\frac{dz}{e^{az}\sqrt{1-\beta^2e^{-2be^{az}}}},
     \label{nullt}
     \en
      where $\beta=P_1/E$ is a constant which is smaller than 1, and
      \be
      x=\int\frac{dz}{e^{az+2be^{az}}\sqrt{1-\beta^2e^{-2be^{az}}}}.
      \label{nullx}
      \en
     Here we directly write the null geodesic by using the coordinates rather than $\tau$, for now
     $\tau$ has no significant physical sense. Our solution
     (\ref{metric}) can be treated as a generalization of
     (\ref{ads}), thus we expect the geodesics will degenerate to that of AdS in the limit $b\to 0$.
     We have shown that the motion equation along $z$ direction is
     just the same as that of AdS, since $b$ does not appear in that
     equation. Now we expand (\ref{nullt}) and (\ref{nullx}) into
     series about $b=0$, respectively,
     \be
     t+c_5=-\frac{e^{-az}}{a({1-\beta^2})^{1/2}}- \frac{\beta^2
     z}{(1-\beta)^{3/2}}b+\frac{\beta^2(2+\beta^2)e^{az}}{2a(1-\beta^2)^{5/2}}b^2
     +{\cal O}[b^3],
     \label{expant}
     \en
     and
    \be
    \beta^{-1}x+c_6=-\frac{e^{-az}}{a(1-\beta^2)^{1/2}}-\frac{(2-\beta^2)z}{(1-\beta^2)^{3/2}}b
    +\frac{(4-2\beta^2+\beta^4)e^{az}}{2a(1-\beta^2)^{5/2}}b^2+ {\cal O}
    [b^3],
    \label{expanx}
      \en
      where $c_5$ and $c_6$ are two integration constants. The
      leading terms in (\ref{expant}, \ref{expanx}) are the
      corresponding terms in the case of AdS. The higher order terms
      denote the corrections to AdS.

 \section{the essence of the source matter}
  Although we get the exact form of $\rho$ and $p$ in (\ref{rho}) and
  (\ref{p}), we do not know what it really is. We will investigate some
  different candidates for the source matters.

  First, it can not be pure dust or photon, since $p\neq 0$ and
  $p\neq \frac{1}{3}\rho$.

  Next, we consider the source matters are consisting of dust and photon. In this case,
  \be
  \rho=\rho_{\rm dust}+\rho_{\rm photon},
  \label{rhodr}
  \en
 \be
  p=p_{\rm photon}.
  \label{pp}
  \en
 $\rho_{\rm photon}$ and $p_{\rm photon}$ are related by
 \be
 p_{\rm photon}=\frac{1}{3} \rho_{\rm photon}.
 \en
  By using (\ref{rho}) and (\ref{p}), we derive
  \be
  p_{\rm photon}=a^2(3+4be^{az}+b^2e^{2az}),
  \en
    \be
   \rho_{\rm photon}=3a^2(3+4be^{az}+b^2e^{2az}),
   \en
 and
 \be
 \rho_{\rm dust}=-6a^2(2+\frac{10}{3}be^{az}+b^2e^{2az}).
 \en
  For
  the special configuration which can perfectly match to Taub space,
   $b=-3e^{1/3},  az_0=-1/3$, $\rho_{\rm dust}$ becomes,
   \be
   \rho_{\rm dust}=-12a^2\left[
   (1-\frac{9}{4}e^{\frac{1}{3}(1-z)})^2 +\frac{1}{16}
   e^{\frac{1}{3}(1-z)} \right]<0.
   \en
   Thus, we find an unusual property of the source matter:
   $\rho_{\rm dust}$ is always negative.  Though it can be realized
  by some quantum effects \cite{bd}, in classical level, we may need to seek more
  natural matter for the source.

  Third, we prove that the source is not a scalar field. In coordinates (\ref{metric}) the
  components of the energy-momentum of a scalar read,
  \be
  T_t^t=\frac{1}{2}g^{tt}\p _t\phi \p_t\phi-\frac{1}{2}( g^{zz}\p _z\phi
  \p_z\phi + g^{xx}\p _x\phi \p_x\phi+g^{yy}\p _y\phi
  \p_y\phi)-V(\phi),
  \label{ttt}
  \en
    \be
  T_z^z=-\frac{1}{2}g^{tt}\p _t\phi \p_t\phi+\frac{1}{2} g^{zz}\p _z\phi
  \p_z\phi - \frac{1}{2}g^{xx}\p _x\phi \p_x\phi-\frac{1}{2}g^{yy}\p _y\phi
  \p_y\phi-V(\phi),
  \label{tzz}
  \en
   \be
  T_x^x=-\frac{1}{2}g^{tt}\p _t\phi \p_t\phi-\frac{1}{2} g^{zz}\p _z\phi
  \p_z\phi + \frac{1}{2}g^{xx}\p _x\phi \p_x\phi-\frac{1}{2}g^{yy}\p _y\phi
  \p_y\phi-V(\phi),
  \label{txx}
  \en
   \be
  T_y^y=-\frac{1}{2}g^{tt}\p _t\phi \p_t\phi-\frac{1}{2} g^{zz}\p _z\phi
  \p_z\phi - \frac{1}{2}g^{xx}\p _x\phi \p_x\phi+\frac{1}{2}g^{yy}\p _y\phi
  \p_y\phi-V(\phi),
  \label{tyy}
  \en
   where $V(\phi)$ denotes the potential of the scalar field.
   Einstein equation requires
   \be
    \rho=-T_t^t,
    \label{energy}
    \en
    \be
    p=T_z^z=T_x^x=T_y^y.
    \label{pressure}
    \en
   (\ref{pressure}) yields,
   \be
   \p_z \phi=e^{-(az+be^{az})}\p_x\phi,
   \label{partial1}
   \en
   and
   \be
   \p_z \phi=e^{-(az+be^{az})}\p_y\phi.
   \label{partial2}
   \en
  From now on we only consider the static solution. Under this
  assumption, we derive the general solution of (\ref{partial1}) and
  (\ref{partial2}),
  \be
  \tilde{\phi}={C}\left[e^{-(az+be^{az})}-a(x+y)+bEi(-be^{az})\right],
  \label{phi}
  \en
  where $C$ is an integration constant, and $Ei$ denotes the exponential integral
  function, which is defined as
  \be
  Ei(u)\triangleq -\int_{-u}^{\infty} ds \frac{~~e^{-s}}{s}.
  \en
  From (\ref{energy}) and (\ref{pressure}), we obtain another
  equation that $\phi$ must satisfy,
  \be
  (\p _z\phi)^2=-4ba^2e^{az}-2b^2a^2e^{2az}.
  \label{p+rho}
  \en
  However, $\tilde{\phi}$ does not satisfy the above equation (\ref{p+rho}). To show
  this, we define
  \be
  F\triangleq (\p _z\tilde{\phi})^2+4ba^2e^{az}+2b^2a^2e^{2az}.
  \en
  Direct calculation presents,
  \be
  F=a^2C^2e^{-2(az+be^{az})}+4ba^2e^{az}+2b^2a^2e^{2az}.
  \en
  We see that $F$ depends on $z$ and hence cannot be identical to
  zero. So it is impossible that a  scalar plays the role of
  the source matter.

   Observing the energy momentum of metric (\ref{metric}) carefully, we find that the
   energy density is prone to be a negative number. Especially in the case of matching a Taub solution
   it is negative. It is natural since this solution is a generalization of AdS.
   Thus, we should consider sources dominated by phantom fields.  The phantom has been widely investigated
  in cosmology after the discovery of cosmic acceleration. It can violate the domination energy condition,
  which  is a requirement of  a reasonable source for
  Taub solution \cite{dec}. And it  just can be realized in metric
  (\ref{metric}) \cite{self}. Thus it would be proper to investigate
  the possibility of a phantom source. The discussions are parallel to the
  case of a scalar.  The energy-momentum for a
  phantom $\psi$ can be written as follows,

   \be
  T_t^t=-\frac{1}{2}g^{tt}\p _t\psi \p_t\psi+\frac{1}{2}( g^{zz}\p _z\psi
  \p_z\psi + g^{xx}\p _x\psi \p_x\psi+g^{yy}\p _y\psi
  \p_y\psi)-U(\psi),
  \label{tttp}
  \en
    \be
  T_z^z=-(-\frac{1}{2}g^{tt}\p _t\psi \p_t\psi+\frac{1}{2} g^{zz}\p _z\psi
  \p_z\psi - \frac{1}{2}g^{xx}\p _x\psi \p_x\psi-\frac{1}{2}g^{yy}\p _y\psi
  \p_y\psi)-U(\psi),
  \label{tzzp}
  \en
   \be
  T_x^x=-(-\frac{1}{2}g^{tt}\p _t\psi \p_t\psi-\frac{1}{2} g^{zz}\p _z\psi
  \p_z\psi + \frac{1}{2}g^{xx}\p _x\psi \p_x\psi-\frac{1}{2}g^{yy}\p _y\psi
  \p_y\psi)-U(\psi),
  \label{txxp}
  \en
   \be
  T_y^y=-(-\frac{1}{2}g^{tt}\p _t\psi \p_t\psi-\frac{1}{2} g^{zz}\p _z\psi
  \p_z\psi - \frac{1}{2}g^{xx}\p _x\psi \p_x\psi+\frac{1}{2}g^{yy}\p _y\psi
  \p_y\psi)-U(\psi),
  \label{tyyp}
  \en
   where $U(\psi)$ denotes the potential of the phantom field.
   Similarly to the case of a scalar, we derive
  \be
  \tilde{\psi}={D}\left[e^{-(az+be^{az})}-a(x+y)+bEi(-be^{az})\right].
  \label{psi}
  \en
   In the context of a phantom,
   (\ref{p+rho}) becomes
   \be
  (\p _z\psi)^2=4ba^2e^{az}+2b^2a^2e^{2az}.
  \label{p+rho1}
  \en
   And defining
    \be
  F\triangleq -(\p _z\tilde{\psi})^2+4ba^2e^{az}+2b^2a^2e^{2az},
    \en
    we find
    \be
  F=-a^2D^2e^{-2(az+be^{az})}+4ba^2e^{az}+2b^2a^2e^{2az},
  \en
  where $D$ is an integration constant.
  Therefore $F$ is a function of $z$ and a single phantom can not
  behave as the source either.

  We see that the source matter may be much complicated than we
   expected. It may be composed by several different components.
   Assuming that every component behaves as perfect fluid, that is,
   it is isotropic,
   \be
   p_{ix}=p_{iy}=p_{iz},
   \en
   where $p_i$ denotes the pressure of the i-th component.
   The constraint equations for $\psi$, similar to that of $\phi$ (\ref{partial1}) and (\ref{partial2}),
    are still valid. The solution
 for general solution for $\psi$  is the same as above,
 \be
  \psi={D}\left[e^{-(az+be^{az})}-a(x+y)+bEi(-be^{az})\right].
  \label{psi1}
  \en
  Now we try to derive the potential $U(\psi)$ through the motion
  equation of the phantom.Our strategy is as follows: One first derives the potential through the motion equation of the
  phantom.
  Second one
  substitutes the potential to the energy momentum form of the
  phantom
  (\ref{tttp}-\ref{tyyp}) to obtain the density of pressure. And then
  one compares the energy momentum form of the metric (\ref{metric})
  with the energy momentum form of the scalar to find which are the
  other necessary components of the source.

   The motion equation of the scalar reads,
   \be
   \frac{1}{\sqrt{-{\rm det}(g)}}\partial _{\mu}\left(g^{\mu\nu}{\sqrt{-{\rm det}(g)}}
   \partial_{\nu} \psi\right)-\frac{dU(\psi)}{d\psi}=0,
   \label{motion}
   \en
   where det$(g)$ marks the determinate of the components of the
   metric (\ref{metric}). Substituting (\ref{psi1}) into above
   equation, we derive
   \be
   U=-a^2D^2\left[(be^{az}-1)e^{-2(az+be^{az})}+2b^2Ei(-2be^{az})\right]+D_1,
   \en
   where $D_1$ is an integration constant. Then, the density and pressure of the
   scalar read,
   \be
   \rho_{\rm phantom}= -\frac{1}{2}a^2C^2e^{-2(az+2be^{az})}
   \left[1+2be^{az}+4b^2e^{2(az+2be^{az})}Ei(-2be^{az})\right]+D_1,
   \en
  \be
  p_{\rm phantom}=\frac{1}{2}a^2C^2e^{-2(az+2be^{az})}
   \left[-1+2be^{az}+4b^2e^{2(az+2be^{az})}Ei(-2be^{az})\right]-D_1.
   \en
   If the source is composed of other components except for the
   phantom, the total density $\rho_{ot}$ and pressure $p_{ot}$ of other components can be
   written as,
   \be
   \rho_{ot}=-a^2(3+8be^{az}+3b^2e^{2az}) +\frac{1}{2}a^2D^2e^{-2(az+2be^{az})}
   \left[1+2be^{az}+4b^2e^{2(az+2be^{az})}Ei(-2be^{az})\right]-D_1,
   \en
   \be
   p_{ot}=a^2(3+4be^{az}+b^2e^{2az})- \frac{1}{2}a^2D^2e^{-2(az+2be^{az})}
   \left[-1+2be^{az}+4b^2e^{2(az+2be^{az})}Ei(-2be^{az})\right]+D_1.
    \en
   We see that the pressure of the components other than the phantom
   does not vanish. Therefore, a single fluid of dust can not play the role.
   The pressure of the other components is not $1/3$ of the density,
   hence a single fluid of photon can not play the role either. We
   consider a source composed of a phantom, dust, and photon. The
   pressure and density  of the photon should be,
   \be
   p_{\rm photon}=a^2(3+4be^{az}+b^2e^{2az})- \frac{1}{2}a^2D^2e^{-2(az+2be^{az})}
   \left[-1+2be^{az}+4b^2e^{2(az+2be^{az})}Ei(-2be^{az})\right]+D_1,
   \en
   and
   \be
   \rho_{\rm photon}=3p_{\rm photon}.
   \en
   The resulting density of the dust reads,
   \be
   \rho_{\rm dust}=-2a^2(6+10be^{az}+3b^2e^{2az})+ a^2D^2e^{-2(az+2be^{az})}
   \left[-1+4be^{az}+8b^2e^{2(az+2be^{az})}Ei(-2be^{az})\right]-4D_1,
     \en
   where the constant $D$ denotes the amplitude of the phantom, and
   $D_1$ yields a effect of a cosmological constant, which can be
   understood as an AdS background.
  For a special configuration which can match Taub solution
  $b=-3e^{-az_0},~a=-1/3,~z_0=1$, it is not difficult to find
  positive definite densities for the dust and photon since we have two constants
  to adjust. For example, $D_1<51.05$ will ensure $\rho_{\rm dust}>0$ in
  the interval $z\in [0,1]$ when $D=1$.

 \section{conclusions and discussions}
   In this letter we explore the geodesic properties and the
   possible substances of the solution presented in \cite{self}.

   We give the exact form of geodesics along $z$ direction, both for
   the time-like and null cases. We find that the global effect of gravitation of the space put forward
    in \cite{self} not only the
    local effect, can be simulated by a field of a uniformly
    accelerating observer.

    We discuss different cases of the source matter: the dust and photon, a scalar, a phantom, a phantom with dust and photon.
    We find that neither a single scalar nor a
    single phantom can play the role of source matter. We find that
    a phantom can sever as the source with dust and photon.

 {\bf Acknowledgments.}
 H.Noh was supported by grant No. C00022 from the Korea Research
 Foundation.

\end{document}